\documentclass{article}
\usepackage{authblk}
\usepackage[utf8]{inputenc}
\usepackage{amsmath}
\usepackage{amssymb}
\usepackage{authblk}
\usepackage{caption}
\usepackage{cite}
\usepackage{xcolor}
\usepackage{color}
\usepackage{float}
\usepackage{geometry}
\usepackage{graphicx}
\usepackage{graphics}
\usepackage{hyperref}
\usepackage{textcomp}
\usepackage{epstopdf}
\usepackage{subcaption}
\usepackage{lscape}
\usepackage{breqn}
\usepackage{soul}

\title{Wahlquist metric revisited}

\author[1]{M. Horta\c{c}su \footnote{E-mail: hortacsu@itu.edu.tr}}
\affil[1]{Mimar Sinan Fine Arts University, Department of Physics, Istanbul, Turkey.}

\begin{document}
	\maketitle

	
	
	
	\begin{abstract}
		\noindent
		Here we continue studying the Wahlquist metric. We know that the wave equation written for a zero mass scalar particle in the background of this metric gives Heun type solutions. To be able to use the existing literature on Heun functions, we try to put our wave equation to the standard form for these functions. Then we calculate the reflection coefficient of a wave coming from infinity and scattered at the center using this formalism using two independent methods

	\end{abstract}

	\section{Introduction}
	\noindent 
	The work on Heun class solutions of wave equations written for particles in the background metric of different metrics is still popular. One of the most recent papers is given in \cite{Dariescu}. Here we study a similar problem for a metric, which was given a while ago.

	\noindent
	The Wahlquist metric \cite{Wahlquist1, wahl2,wahl3} is an exact interior solution for a finite rotating body of perfect fluid 
	to Einstein's field equations with equation of state corresponding to constant gravitational mass density. It is also given in the celebrated book {\it {Exact Solutions to Einstein's Field Equations}} \cite{Schmudzig}.  This metric is axially symmetric, stationary and is  type D. As stated in 
	\cite{birkandan}, quoting Wahlquist \cite{Wahlquist1}, it can be  ``described as a superposition of a Kerr-NUT metric  \cite{Kerr, Newman} and is a rigidly rotation perfect fluid in the same space-time region". Equation of the state for the perfect fluid is $\rho+ 3p= \mu_0$, a constant. 
	
	\noindent
	The original metric written in \cite{Wahlquist1}, was slightly modified by Senovilla \cite{Jose, Jose1}, and put to new form by Mars \cite{Mars}, "to show that the Kerr-de Sitter and Kerr metrics are contained as subcases".
	Mars states in \cite {Mars} that Kramer \cite{Kramer} showed the vanishing of the Simon tensor \cite{Simon}  for this metric. Mars also states that "the space-time admits a Killing tensor as shown in \cite {Papa}". 
	
	\noindent
	A negative point about this metric is the paper,  \cite{Bradley}, where it was shown that "the Wahlquist perfect fluid space-time can not be smoothly joined to an exterior asymptotically flat vacuum region since the conditions for matching the induced metrics and extrinsic curvatures are mutually contradictory". Thus, " the Wahlquist metric can not describe an isolated rotating body". Bradley et al., however, showed that this obstruction is removed for a limiting form of the Wahlquist solution \cite{Bradley1}, and also for every D-type metric \cite{bradley2}, if the requirement of an asymptotically flatness is removed.
	
	\noindent
	More recent work in this field exists, "where the existence of a rank-2 generalized closed conformal Killing-Yano tensor with a skew-symmetric torsion " \cite{Houri1}, and "the separability of the Maxwell equation on the Wahlquist spacetime" are shown  \cite{Houri2}. A special feature of this metric is that in the wave equation for a massless scalar particle in the background of this metric gives trivial exponential solutions for two of the coordinates, the differential equation for the angular and the radial coordinates may be arranged to give exactly the same equation yielding general Heun solutions. Then the equation is reduced to the Mathieu equation by putting many of the constants in the original equation to zero.  This makes it possible to calculate the two point function in one less dimension \cite{birkandan}.

	\noindent 
	Here we start with the metric as given in \cite{Houri1} and  try to write the wave equation, in the background of this metric, as given in \cite{birkandan}, in the standard form given by \cite{Arscott}. How to perform this task is described in \cite{Forsyth}, as quoted by \cite{Arscott1}, and used meticulously by \cite{Suzuki, Suzuki1}. The same method is recently used in \cite{Vieira, Hatsuda, Noda1, Noda2}. We need the standard form of the wave equation to be able to apply the information given in the existing literature to our work.
	
	\noindent 
	We will first summarize our previous work, \cite{birkandan}. Then we show how to convert the wave equation to the standard form. In the third section, we give the approximate reflection coefficient if a wave, coming from infinity is scattered at the origin. We conclude with a few remarks.
	
	Our aim to revise our work is to show that, as also stated in the first form of this paper, to choose another method in the calculation of the reflection coefficients. The method we used to calculate  these coefficients in the first version of this paper, published within the conference proceedings by IOP, in  J.Phys.Conf.Ser. 2191 (2022) 1, 012015, contribution to: DERELİ-FS-2021, may have some questinable parts, like  the reflection coefficient  being a  ratio of two infinite numbers •  Since it was published in a separate form, we could not publish an Addendum or Correction to this paper. 
	
	\section {Summary of former work}
	Here, we will summarize the work in \cite{birkandan} below.  
	First,  using the metric given in \cite{ Wahlquist1}, 
	we try to calculate $\phi$ where it obeys the equation 
	\begin{equation}
		\frac{1}{\sqrt{g}}\partial_{\mu} g^{\mu \nu} \sqrt{g}  \partial_{\nu} \phi  = 0.
	\end{equation}
	Here $g$ is the determinant of the metric coefficients $g_{\mu \nu}$.
	
	\noindent
	We first write the metric as is given in \cite{Houri1}, which is equivalent to the one given in \cite{Mars}. We follow some of the work in \cite {Houri1} in the first part, and \cite{birkandan} in the second part of this section.
	
	\noindent
	Here "the comoving, pseudoconfocal, spatial coordinates" are used, which are closely related to the oblate-spheroidal coordinates in Euclidean geometry" \cite{Wahlquist1}. We set $4\pi G= c=1$. Then, the metric reads
	\begin{equation}
		ds^2= (v_{1} +v_{2}) \big(\frac{dz^2}{U}+ \frac{dw^2}{V} \big) 
		+\frac{U}{v_{1} +v_{2}}(d\tau+v_{2}d\sigma)^2-\frac{V}{v_{1} +v_{2}}
		(d\tau-v_{1}d\sigma)^2 
	\end{equation}
	where
	\begin{eqnarray}
		&& U=Q_0+a_1{\frac{sinh(2\beta z)}{2\beta}} 
		-\frac{\nu_0}{\beta^2}{\frac{cosh(2\beta z)-1}{2\beta^2}}
		\nonumber \\&&
		-{\frac{\mu_0}{2\beta^2}}\big[{\frac{cosh(2\beta z)-1}{2\beta^2}} -z{\frac{sinh(2\beta z)}{2\beta}}\big],
	\end{eqnarray}
	\begin{eqnarray}
		&& V=Q_0+a_2{\frac{sin(2\beta w)}{2\beta}} 
		+\frac{\nu_0}{\beta^2}{\frac{-cos(2\beta w)+1}{2\beta^2}}
		\nonumber \\&&
		-{\frac{\mu_0}{2\beta^2}}\big[{\frac{cos(2\beta w)-1}{2\beta^2}} +w{\frac{sin(2\beta w)}{2\beta}}\big]
	\end{eqnarray}
	where
	\begin{equation}
		v_1=\frac{cosh(2\beta z)-1}{2\beta^2},
		v_2=\frac{-cos(2\beta w)+1}{2\beta^2}.
	\end{equation}
	This metric has six real constants $Q_0,a_1,a_2,\nu_0,\mu_0,\beta$.
	Here $a_1$ is related to the NUT \cite{Newman1,Halilsoy} parameter; $a_2$ is related to the mass parameter. One writes the other variables to express the energy density, pressure and the fluid velocity of the perfect fluid. 
	Both the pressure, $p=\frac{1}{2} \mu_0(1-\kappa^2 f)$, and the density of the Wahlquist fluid  $\mu$, 
	are given \cite{Wahlquist1} in terms of $\mu_0$ and metric coefficients and obey $\mu+3p=\mu_{0}.$ 
	Here $\beta$ is related to the scaling of $z$ and $w$, both space coordinates, in the linear transformation of these variables that are given in the original paper by Wahlquist \cite{Wahlquist1}. $U$ and $V$ are related to $h_2,h_1$ in the original metric. All these parameters are scaled so that they do not vanish in the $\beta$ going to zero limit.
	
	\noindent
	When we use the ansatz
	\begin{equation}
		\phi = R(z) Y(w) T(\tau) S(\sigma)
	\end{equation}
	for the solution, the wave equation written in this metric separates easily. We have two Killing vectors since the metric does not depend on  $\tau$ and $\sigma$, related to $t$ and $\theta$ in the original metric \cite{Wahlquist1}. If we make a Wick rotation which changes $w$ to $y=iw$, $a_2$ to $-ia_2$ \cite{Mars} where $i$ is the square root of minus unity, \cite{Mars},
	the metric becomes symmetrical
	\begin{equation}
		ds^2= (v_{1} +v_{2}) \big(\frac{dz^2}{U}- \frac{dy^2}{V} \big) 
		+\frac{U}{v_{1} +v_{2}}(d\tau-v_{2}d\sigma)^2-\frac{V}{v_{1} +v_{2}}
		(d\tau+v_{1}d\sigma)^2.
	\end{equation}
	Then we have identical equations for $z$ and $y$. The equation for $z$ reads
	\begin{equation}
		\partial_{z}(U \partial_{z}) R(z) + \big(\frac{v_1^2}{U}\partial_{\tau}^{2}+2\frac{v_1 }{U}\partial_{\tau}\partial_{\sigma}+\frac{1}{U} \partial_{\sigma}^{2}\big) R+\Lambda R=0.
	\end{equation}
	We get exactly the same equation for the new variable $y$, with appropriate changes like $U$ going to $V$  and $v_1$ going to $v_2$.
	\begin{equation}
		\partial_{y}(V \partial_{y}) S(y) + \big(\frac{v_2^2}{V}\partial_{\tau}^{2}+2\frac{v_2 }{V}\partial_{\tau}\partial_{\sigma}+\frac{1}{V} 
		\partial_{\sigma}^{2}\big) S(y)=\Lambda S(y).
	\end{equation}
	Since the functions $v_1, U$ ( similarly $v_2, V$) don't depend on $\tau$ and $\sigma$, the solutions for $\tau$ and $\sigma$ are just exponential functions, giving us constants upon differentiation.
	
	\noindent
	Here we will first try to write the differential equation for a zero mass scalar field for the radial coordinate $z$, in the background of this metric. Then, we make more independent variable changes to put our equation in the standard form, as given in \cite{Arscott}.
	
	\noindent 
	We find that this is not an easy task. The presence of hyperbolic sine and cosine functions in the wave equation prevents us from using standard analytical methods in solving differential equations. We change our variables to $x=exp(2\beta z)$ which makes it possible to write the hyperbolic sine and cosine functions in terms of powers.
	$sinh {2\beta z} = 1/2[x-1/x]$, $ cosh{2\beta  z}-1=1/2[x+1/x-2]$.
	Then, however, exists the relation  $z=ln{x}/(2\beta)$. 
	
	\noindent
	The codes we use to solve differential equations analytically do not recognize  $ln{x}$. Thus, we can not give an exact solution for all values of the independent variable $x$. We can get solutions only in one patch, by using an algebraic expression approximating $lnx$ in this region. 
	
	\noindent
	Upon the variable change from $z$ to  $x$, we get the new equation
	\begin{equation}
		\frac {d^2 \psi(x)}{dx^2}
		+\big(\frac{{\frac{d U}{d x }}}{U} +\frac{1}{x}\big) \frac{d }{d x } \psi(x) - \frac{1}{x^2 U^2}\big(\omega^2v_1^2+2v_1\omega s +s^2+\Lambda\big) \psi(x)=0.
	\end{equation}
	Here we used the solution to the $\tau$ equation in the form $exp(-i\omega \tau)$ and the solution for the $i\sigma$ equation a solution in the form $exp(-i\sigma s)$. Note that our metric had Killing vectors for both $\tau$ and $\sigma$.
	Here
	\begin{equation} 
		U=Q_0 + a_1\frac{(x-1}{4x\beta}-\big[\frac{\mu_0}{4\beta^4}+\frac{\nu_0}{4\beta^3}]\frac{(x-1)^2}{x}+ \frac{\mu_0}{8\beta^4} \frac{(x^2-1)}{x} ln x\big],
	\end{equation}
	\begin{equation}
		v_1= \frac{(x-1)^2}{4\beta^2 x},    
	\end{equation}
	\begin{equation}
		\frac{dU}{dx}= \big[\frac{a_1}{\beta}(1+\frac{1}{x^2})-
		(\frac{\nu_0}{\beta^2} + \frac{\mu_{0}}{\beta^4}(1-\frac{1}{x^2})+ \frac{\mu_0}{8\beta^4 x }(x-\frac{1}{x}) +\frac{\mu_0}{8\beta^4} \ln{x} (1+\frac{1}{x^2})\big] .
	\end{equation}
	From here on, we will define $\mu_0'=\frac{\mu_0}{(8\beta^4)}$ and $\nu_0'= \frac{\nu_0}{(4\beta^3)}$, $a'_1= a_1/\beta$ and use these new constants in our equations.
	
	\noindent
	By studying the equation written in terms of the variable $x$, we know that there is a singularity near $x=1$.  To study the solution around this singularity, we expand $lnx$ in the neighborhood of this point and use $x-1$ instead of $lnx$ in the wave equation we use in our calculation, after we differentiate $U$ with respect to $x$.
	
	\noindent
	We find that, if we keep all our constants non-zero, we get \cite{birkandan} the exact solution in terms of the general Heun function \cite {Arscott, Slavyanov, Fiziev, Hortacsu}  up to an exponential and terms (powers of polynomials) multiplying this function. The regular singularities are at zero and at three other finite points and at infinity,.

	\noindent
	Then we try the next possible choice.
	We try to find the solution when we keep all the terms in $U$ aside from $Q_0$, which we set equal to zero. Then, our equation,  again, has five regular singularities at $0,1$, infinity, and at the points 
	\begin{equation}
		x_a= 
		-\frac{1}{2\mu'_0}(M - [N]^{1/2}),   
	\end{equation}
	\begin{equation}
		x_b= 
		-\frac{1}{2\mu'_0}(M + [N]^{1/2}),   
	\end{equation}
	where
	\begin{equation}
		M=(a'_1-\nu'_0-2\mu'_0),
	\end{equation}
	\begin{equation}
		N^2=M^2 - 4\mu'_0(a'_1+\nu'_0+\mu'_0). 
	\end{equation}
	
	\noindent
	The  $HeunG$,  General Heun form solution is retained when we set $Q_0$ to zero.
	To further simplify our expressions, we define $a''_1= \frac{a'_1}{\mu'_0},\nu''_0= \frac{\nu'_0}{\mu'_0} $.

	\section{Reduction to the standard form}
	
	How to reduce an equation with four regular singular points at finite values to an equation with singularities at zero, unity, a finite point, and infinity is described in \cite{Arscott1}. If the coefficients of the first derivatives satisfy a linear relation, one just has to make a homographic substitution to bring three of the original singularities  $b_1, b_2, b_4$ to zero,  unity, and infinity.  The previous regular point at infinity is transformed to $1-x_b$. This point is decoupled from the equation as shown by \cite{Suzuki}.   This transformation is 
	\begin{equation}
		u= \frac{x-b_1}{x-b_3} \frac{b_2-b_3}{b_2-b_1}.
	\end{equation}
	Then 
	in our case $b_1=0, b_1=1,b_3=x_b, b_4=x_a  $, which gives $u=\frac{ x (1-x_b)}{x-x_b}$. The singularity at $x_b$ is moved to infinity, the singularity at $x_a$ is moved to $ \frac{-x_a(1-x_b)}{x_b-x_a}$.
	Then our equation reads
	\begin{eqnarray}
		&&
		\frac{d^2 \phi(u)}{du^2} +\big( \frac{1}{x_a x_b u}-\frac{1}{u-1} +
		\frac{(1+x_b)-x_a x_b \frac{1}{x_a}} {(1-x_a)(x_b-x_a)(u+\frac{x_a(1-x_b)}{x_b-x_a})} \big) 
		\frac{d \phi(u)}{du} + \frac{J}{2 Tu} \phi(u)
		\nonumber \\ &&
		+ \frac{1}{4}\big( \frac{D}{T^2} + \frac {2s\omega}{E u T }+ \frac {2 s^2}{F (u-1) T} +\frac{4s^2}{H(u-1)^2}+\frac{\omega^2}{4(x_a x_bu)^2}+\Lambda\big) \phi(u)=0.
	\end{eqnarray}
	\begin{equation}
		T=  (u+\frac{x_a(1-x_b)}{x_b-x_a}),
	\end{equation}
	\begin{equation}
		D= \frac{1}{(x_b-x_a)^2} (\frac{\omega(1-x_a)}{2\beta^2x_a}- \frac{2s}{1-x_a})^2,  
	\end{equation}
	\begin{equation}
		E= 4x_a x_b (x_b-x_a),
	\end{equation}
	\begin{equation}
		F=(1-x_a)^2 (1-x_b)(x_b-x_a)
	\end{equation}
	\begin{equation}
		H= (1-x_a)^2 (1-x_b)^2,
	\end{equation}
	\begin{equation}
		J=- \frac {\omega^2 x_b(1-x_a)}{\beta^4 x_a^2 x_b^2 (x_b-x_a)}
	\end{equation}
	
	\noindent
	The new equation is not in the standard General Heun form, though. One has to remove the terms where the singularities are in quadratic form from the new differential equation. One makes an F-homotopic transformation on the dependent variable $\phi(u)$ in the form
	\begin{equation}
		\phi(u) = (u)^{\lambda} (u-1)^{\xi}
		(u+\frac{x_a(1-x_b)}{x_b-x_a})^{\kappa} g(u),
	\end{equation}
	which gives rather complicated expressions for these indices.
	\begin{equation}
		\lambda= \frac{1}{2}[1\pm{\sqrt{1-\frac{w^2}{4\beta^4 (x_a x_b)^2}}}],
	\end{equation}
	
	\begin{equation}
		\xi=1\pm{\sqrt{1- \frac{4s^2}{(1-x_a)^2 (1-x_b)2}}},
	\end{equation}
	\begin{equation}
		\kappa = -\frac{1}{2} [\frac{x_a}{x_a-x_b}-1 \pm{\sqrt{(\frac{x_a}{x_a-x_b})^2-
				(\frac{w(x_a-1)}{x_a\beta^2}- \frac{4s}{(x_a-1) })^2}}].
	\end{equation}
	
	\noindent
	Now we can write the new wave equation as
	\begin{equation}
		\frac{d^2 g(u)}{du^2}
		+\big(\frac{ c}{u} + \frac {d}{u-1} + \frac{e}{(u+\frac{x_a(1-x_b)}{x_b-x_a})} \big) \frac{d }{d u } g(u) + \frac {  A B u + Q+\Lambda}{u(u-1) (u+\frac{x_a(1-x_b)}{x_b-x_a})}g(u)=0.
	\end{equation}
	Here
	\begin{equation}
		c=\big[1-{\sqrt{1-\frac{w^2}{4\beta^4 (x_a x_b)^2}}}\big],
	\end{equation}
	\begin{equation}
		d = 1 - 2{\sqrt{1- \frac{s^2}{(1-x_a)^2 (1-x_b)2}}},
	\end{equation}
	\begin{equation}
		e= 1- {\sqrt{{(\frac{x_a}{x_a-x_b})^2-
					(\frac{w(x_a-1)}{x_a\beta^2}- \frac{4s}{(x_a-1) })^2}}}
	\end{equation}
	Using the relation
	\begin{equation}
		c+ d +e = A +B +1 ,   
	\end{equation}
	which is valid when our equation has Heun form,
	we can also calculate $A$ and $B$.
	\begin{equation}
		A= \frac{1}{2} ( c + d + e -1 + \sqrt{( c+d + e -1)^2 -4 S } )  ,
	\end{equation}
	\begin{equation}
		B=\frac{1}{2} ( c+ d+ e -1 - \sqrt{( c+d + e -1)^2 -4 S } ).  
	\end{equation}
	The expression inside the square root above simplifies
	\begin{eqnarray}
		&&
		( c + d + e -1)^2 -4 S =
		[\frac{2}{x_a x_b} + (-1 + \frac{ x_a x_b -1}{x_b(x_b-x_a)})^2
		\nonumber \\ &&
		-[\frac{\omega(1-x_b)}{2 \beta^2 x_b(x_b-x_a)}-\frac{2s}{(1-x_b)(x_b-x_a)} ]^2].
	\end{eqnarray}
	Here we did not give S explicitly, since it is given in the equation given above, and does not appear anymore.
	
	We  also give
	\begin{equation}
		Q_1=-\frac{1}{2} [e c + \frac{x_a(1-x_b)d c} {(x_b-x_a)} + \frac{x_a(1-x_b)}{2x_a x_b(x_b-x_a)}+ \frac{1+x_b-x_a x_b-\frac{1}{x_a}}{2(x_b-x_a)(1-x_a)x_a x_b}],
	\end{equation}
	\begin{equation}
		Q_2= \frac{\omega^2 x_b(1-x_a)}{2\beta^4 x_a^2 x_b^2 (x_b-x_a)} - \frac{s\omega}{2\beta^2 x_a x_b (x_b-x_a)}.
	\end{equation}
	In our expression for the wave equation and the solution, we then use $Q$ where $Q=Q_1+Q_2+\Lambda$,
	
	\section{ Scattering at the origin}
	\noindent
	While changing our independent variable from $z$ to $x$, we overlooked one point. Now $x$ corresponds to minus infinity for $z$. To scatter from origin, we need to take $x=1$. We, therefore, change our variables to $u-1=-y$. We know that $u=1$ is the same point when $x=1$. Then our equation reads
	\begin{equation}
		\frac{d^2 g(y)}{dy^2}
		+\big(\frac{ c}{y-1} + \frac {d}{y} + \frac{e}{(y-\frac{x_b(1-x_a)}{x_b-x_a})} \big) \frac{d }{d y } g(y) + \frac {A B (y-1) - Q+\Lambda}{y(y-1) (y-\frac{x_b(1-x_a)}{x_b-x_a})}g(y)=0,
	\end{equation}
	which gives as solution
	\begin{equation}
		g(y)= H_G(\frac {x_b(1-x_a)}{x_b-x_a}, -Q-A B ; A,  B, d, c, e; y=1-u) ,  
	\end{equation}
	where $H_G$ is the generalized Heun function, in the standard form\cite{Arscott}.
	We had a singular point at infinity. We want to have our wave to come from infinity and scatter at the origin. The other singular point at $x_a$ can be arranged to be at a complex value.
	
	\noindent
	For the scattering process, we will first use the formula given by Dekar et al \cite {Dekar}.  This formula is between two finite points. We bring the point infinity to unity by using the transformation $t= \frac{y}{y-1}$. Now the point where $y$ and $t$ are equal to zero coincide, and $y$ going to infinity is given by $t=1$. 
	
	The problem with this calculation is that one gets the reflection coefficients in terms of Heun functions at one of its singular points, namely unity. 
	
	\noindent
	The new equation is not in the Heun form. To bring it back to the Heun form, we multiply the solution $g((t)$ by a power
	\begin{equation}
		g(t) = (t-1)^{A} h(t)
	\end{equation}
	and try to use $h(t)$ in our further calculations.
	Our new dependent variable $h(t)$ satisfies the equation
	\begin{equation}
		\frac{d^2 h(t)}{dt^2}
		+\big(\frac{ A +1- B}{t-1} + \frac{ d }{t}
		+\frac{ e}{t-\frac{x_b(1-x_a)}{x_a(1-x_b)}}\big)
		\frac{d}{dt} h(t)+
		\big(\frac{( A(d + e - B)t+Q_5+\Lambda}
		{t(t-1)(t-\frac{x_b(1-x_a)}{x_a(1-x_b)})}\big)h(t)=0.
	\end{equation}
	\noindent
	At this point, we make a change in our notation. We want to use the standard Heun notation. This means we will redefine our parameters as $A+1-B=\delta,  d=\gamma, d+ e -B = \beta$.  The other parameters, defined as $ e=\epsilon, A=\alpha$ will remain as they are. Now, we can use the standard notation, since our equations will be of the Heun form in the further part of this work, and we will be able to use known relations for this function.
	
	\noindent
	Now, the differential equation is written as
	\begin{equation}
		\frac{d^2 h(t)}{dt^2}
		+\big(\frac{ \gamma}{t} + \frac{\delta}{t-1}
		+\frac{\epsilon}{t-\frac{x_b(1-x_a)}{x_a(1-x_b)}}\big)
		\frac{d}{dt} h(t)+
		\big(\frac{(\alpha \beta)t+Q_5+\Lambda}
		{t(t-1)(t-\frac{x_b(1-x_a)}{x_a(1-x_b)})}\big)h(t)=0.
	\end{equation}
	\begin{equation}
		Q_5=\frac{(x_b-x_a)( \alpha (\gamma + \epsilon- \beta)-Q)-\gamma( \alpha 
			x_b(1-x_a)}{x_a(1-x_b)}.
	\end{equation}
	From now on, we will rename $h(t)$ as $y_1$.
	
\begin{figure}[H]
		\begin{subfigure}{.5\textwidth}
			\centering
			\includegraphics[width=0.9\linewidth]{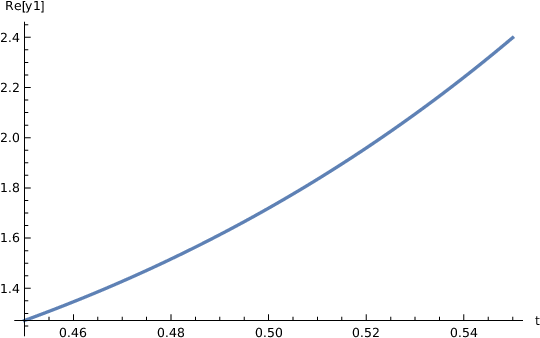}
			\caption{Real part of $y_1$.}
			\label{fig:sub-firstang0}
		\end{subfigure}
		\begin{subfigure}{.5\textwidth}
			\centering
			\includegraphics[width=0.9\linewidth]{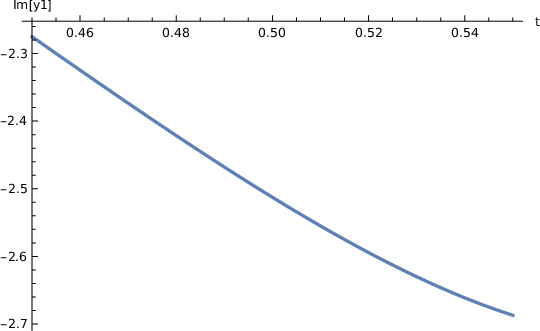}
			\caption{Imaginary part of $y_1$.}
			\label{fig:sub-secondang0}
		\end{subfigure}
		\caption{Plot of $y_1$.}
		\label{fig:y1}
	\end{figure}

	Figure (\ref{fig:y1}) shows $y_1$ around $t=0.5$. Other parameters are taken as $x_a = 4, x_b = 3,s = 10, \omega = 1,\Lambda = 1,\beta_1 = 1000$ in all plots of this paper.
	Now we have to write  $y_1$ in terms of the two linearly independent solutions, which we name $y_3(1-t)$ and $y_4(1-t)$, the two linearly independent solutions one gets when one expands around $1-t=v$.
	\noindent
	In other words, we want to write 
	\begin{equation}
		y_1(t)= D_1 y_3 (1-t) + D_2 y_4(1-t).
	\end{equation}
	The right-hand side gives us the two Heun solutions when the same equation is expanded as a function of $1-t$.
	This expansion is easily done
	\begin{equation}
		y_3= H_G \big( \frac{(x_a-x_b)}{x_a(1-x_b)}, Q_3; \alpha,\beta;  \delta,\gamma,\epsilon; v=1-t \big), 
	\end{equation}
	where $Q_3=Q_5-\alpha \beta$, and
	which is the solution of the equation
	\begin{equation}
		\frac{d^2 y_3 (v)}{dv^2}
		+\big(\frac{ \gamma}{v-1} + \frac{\delta}{v}
		+\frac{\epsilon}{v-\frac{x_a-x_b}{x_a(1-x_b)}}\big)
		\frac{d}{dv} y_3(v)+
		\big(\frac{(\alpha \beta)v+Q_3+\Lambda}
		{v(v-1)(v-\frac{(x_a-x_b)}{x_a(1-x_b)})}\big)y_3(v)=0.
	\end{equation}
	\begin{figure}[H]
		\begin{subfigure}{.5\textwidth}
			\centering
			\includegraphics[width=0.9\linewidth]{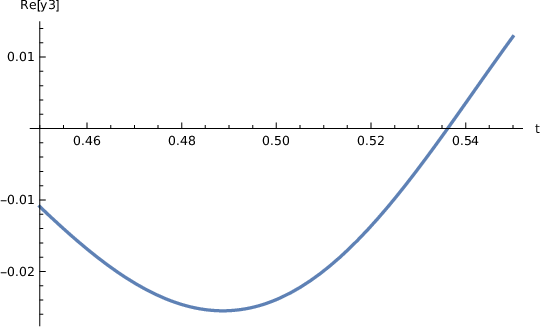}
			\caption{Real part of $y_3$.}
			\label{fig:sub-firstang0}
		\end{subfigure}
		\begin{subfigure}{.5\textwidth}
			\centering
			\includegraphics[width=0.9\linewidth]{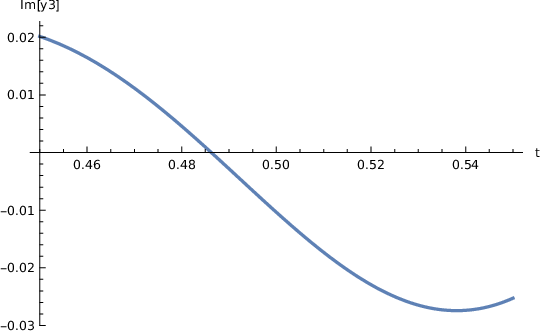}
			\caption{Imaginary part of $y_3$.}
			\label{fig:sub-secondang0}
		\end{subfigure}
		\caption{Plot of $y_3$.}
		\label{fig:DCHcompare}
	\end{figure}
	Then, we make the transformation
	\begin{equation}
		y_4(v) =  v^{(1-\delta)} j(v), 
	\end{equation}
	where $j(v)$ satisfies the equation
	\begin{equation}
		\frac{d^2 j(v)}{dv^2}
		+\big(\frac{ \gamma}{v-1} + \frac{2-\delta}{v}
		+\frac{\epsilon}{v-\frac{x_a-x_b}{x_a(1-x_b)}}\big)
		\frac{d}{dv} j(v)+
		\big(\frac{(\gamma+\epsilon-\alpha) (\gamma+\epsilon-\beta)v+Q_4+\Lambda}
		{v(v-1)(v-\frac{(x_a-x_b)}{x_a(1-x_b)})}\big)j(v) =0.
	\end{equation}
	to give
	\begin{equation}
		y_4= (1-t)^{(1-\delta)}(H_G (\frac{x_a-x_b)}{x_a(1-x_b)},Q_4;  \gamma+ \epsilon- \alpha), \gamma+ \epsilon- \beta; 2-\delta, \gamma,\epsilon; v=1-t),
	\end{equation}
	where $Q_4=Q_3-(1-\delta)(\gamma \frac{x_a-x_b}{x_a(1-x_b)}+ \epsilon)$.
	Note that this last transformation is one of the transformations which does not change the Heun form of the differential equation. 
	Furthermore, the relation $ 1+\alpha+\beta= \gamma+\delta+ \epsilon  $ is valid in this part of our calculations.
	\begin{figure}[H]
		\begin{subfigure}{.5\textwidth}
			\centering
			\includegraphics[width=0.9\linewidth]{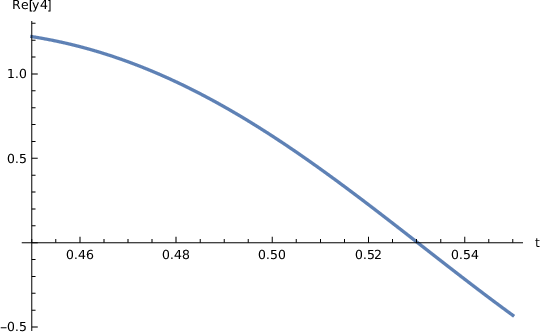}
			\caption{Real part of $y_4$.}
			\label{fig:sub-firstang0}
		\end{subfigure}
		\begin{subfigure}{.5\textwidth}
			\centering
			\includegraphics[width=0.9\linewidth]{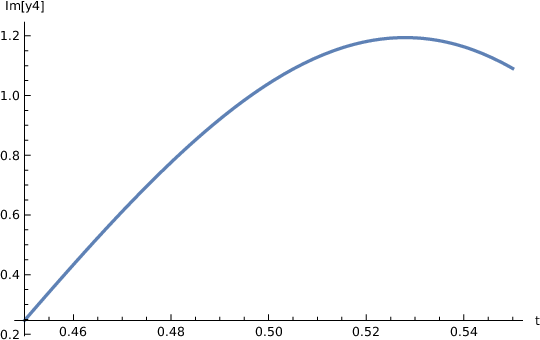}
			\caption{Imaginary part of $y_4$.}
			\label{fig:sub-secondang0}
		\end{subfigure}
		\caption{Plot of $y_4$.}
		\label{fig:DCHcompare}
	\end{figure}
	
	\noindent
	At this point we use the formula given in \cite{Dekar} for expanding $H_G(t)$ in terms of two linearly independent solutions of the equation written for $H_G(1-t=v)$, our equation (46).
	\begin{equation}
		y_1(t)= D_1 y_3 (1-t) + D_2 y_4(1-t).
	\end{equation}
	We find 
	\begin{equation}
		D_1= H_G(\frac{x_b(1-x_a)}{x_a(1-x_b)}, Q_5;\alpha,\beta;\gamma,\epsilon;1),
	\end{equation}
	\begin{equation}
		C_2=H_G\Big( \frac{x_b(1-x_a)}{x_a(1-x_b)},Q_5+( \frac{x_b(1-x_a)}{x_a(1-x_b)}) \gamma(1-\epsilon); \gamma+\epsilon-\alpha, \gamma+\epsilon-\beta;\gamma,\epsilon;1 \Big).
	\end{equation}
	Note that the factor multiplying $j(1-t)$ in $y_4$ can be written as 
	\begin{equation}
		\exp[{(2i{\sqrt{-1+ \frac{s^2}{(1-x_a)^2 (1-x_b)2}}} \ln{1-t}}] . 
	\end{equation}
	Then we can write our eq. (46,52) as
	\begin{eqnarray}
		&&\exp{-i(V \ln(1-t)}) y_1(t) \nonumber \\ &&
		= D_1 {\exp{-i( V \ln(1-t)) }
			y_3(1-t)+D_2 \exp{(iV \ln(1-t)}) j(1-t) },
	\end{eqnarray}
	where
	\begin{equation}
		V= {\sqrt{-1+ \frac{s^2}{(1-x_a)^2 (1-x_b)^2}}}.  
	\end{equation}
	Then our reflection coefficient is given by
	\begin{equation}
		R = \bigg | \frac {D_2}{D_1} \bigg |^2.
	\end{equation}
	This calculation is a formal one. We have to make sure that the Heun function, found by expansion around zero, is also finite at the second singular point for appropriate values of the parameters of our wave equation. A proper Leaver analysis \cite{Leaver}  shows that for $\frac{x_a(1-x_b)}{x_b(1-x_a)}<1$, these terms are convergent when the independent variable equals unity. There is a catch here. Leaver analysis actually gives two roots for convergence for the Heun function at unity. One is, as stated above, $\frac{x_a(1-x_b)}{x_b(1-x_a)}<1$. The other root is unity, for which the convergence when the variable equals unity can not be realized. In writing our connection formula, we could not control which one of these two roots would be taken by our solution. We found that these coefficients were not finite at the point of evaluation.

	A fool proof way to to make sure that these solutions are convergent both at zero and unity, we can try to reduce them to Heun polynomials. It 
	is stated in \cite{Arscott} that to get simple polynomial solutions we have to set $\alpha= -n$. At least we can show that the first few polynomial solutions may be calculated. \cite {Vieira}. The rest can be obtained in the same fashion. Here we have an advantage. We have the same equation both for the radial and the angular solutions. Radial solutions will be sufficient for our purposes here. 
	
	To use this method, we first have to expand  the equation [44] in a power series expansion $ \Sigma_{n=0}^{\infty} c_n x^n $ we get three equations to be satisfied \cite{Arscott, Vieira}
	\begin{eqnarray}
		S_0 c_0+ R_0 c_1=0 \nonumber \\
		P_n C_{n-1} + S_r c_n + R_r c_{n+1} =0 \nonumber \\
		P_n c_{n-1}+ S_r c_n =0, n= 1,2,....., n-1
	\end{eqnarray}
	where
	\begin{eqnarray}
		S_r= -T_r-Q \nonumber \\
		P_r=(n-1+\alpha)(n-1+\beta) \nonumber \\
		T_r = n[(n-1+ \gamma)(1+a) + a\delta) + \epsilon] \nonumber \\
		R_r = (n+1)(n+\gamma) a, a=
	\end{eqnarray}
	We have to make sure that \begin{equation}
		\gamma \ne 0,-1,-2,...
	\end{equation}
	In this method we have to arrange such that the accessory parameter in the Heun equation,  we label as $Q_=Q_5+\Lambda$ as in our equation [44]  will be fixed differently for any power of x in the expansion above. Since the accessory parameter $Q$ includes $\Lambda$, the seperation parameter, this will give a kind of "energy quantization".
	
	As it is done in \cite {Vieira},we can fix $c_0=1$. Then $c_1= \frac{Q_{1,0}}{a\gamma}$, which  gives  the value zero to $Q_{1,0}$ for this solution. 
	
	For the second polynomial, we write 
	\begin{equation}
		c_0+c_1 x= 1 + \frac{Q_{1,1}}{a\gamma} x; a=\frac{x_b(1-x_a)}{x_a(1-x_b)}
	\end{equation}
	Since we have to set $c_2=0$ where, we have to equate
	\begin{equation}
		D_2= \frac{ [\gamma(1+a)+a\delta +\epsilon 
			+ Q_{1,1} ]Q_{1,1} -a \alpha \beta \gamma }{2a^2 \gamma (1+\gamma)}=0
	\end{equation}
	Then
	\begin{equation}
		Q_{1,1}= \frac{[-\gamma(1+a)+a\delta +\epsilon] -\pm \sqrt{T }}{2}
	\end{equation}
	where
	\begin{equation}
		T= [\gamma(1+a)+a\delta +\epsilon]^2 + 4a \alpha\beta \gamma
	\end{equation}
	Then the second term reads 
	\begin{equation}
		H_{P1} 1+ \frac{[-\gamma(1+a)+a\delta +\epsilon] -\pm \sqrt{T}  }{2a \gamma}  x 
	\end{equation}
	We choose the appropriate sign as needed.   
	The other polynomials can be calculated along these lines. Here we used the reference \cite {Vieira} word by word.  Here, we, therefore use a different method to fix our constants in our connection equation, instead of writing the solution in terms of the solutions at unity. This different method was used by \cite {Hatsuda} and Motohashi and Noda \cite {Noda1, Noda2} in their recent papers.
	
	\noindent
	To use this method, we first have to find the second solution of the equation [44]. 
	We set
	\begin{equation}
		h(t) = t^{(1-\gamma)} n(t).  
	\end{equation}
	Then the new equation reads 
	\noindent
	\begin{equation}
		\frac{d^2 n(t)}{dt^2}
		+\big(\frac{2- \gamma}{t} + \frac{\delta}{t-1}
		+\frac{\epsilon}{t-\frac{x_b(1-x_a)}{x_a(1-x_b)}}\big)
		\frac{d}{dt}n(t)+
		\big(\frac{(\alpha +1-\gamma) ( \beta +1- \gamma))t+Q_6+\Lambda}
		{t(t-1)(\frac{t-x_b(1-x_a)}{x_a(1-x_b)})}\big)n(t)=0.
	\end{equation}
	\begin{equation}
		Q_6= Q_5+ (\gamma-1)(\frac{x_b \delta(1-x_a)}{x_a(1-x_b)}+ \epsilon)  
	\end{equation}
	\begin{figure}[H]
		\begin{subfigure}{.5\textwidth}
			\centering
			\includegraphics[width=0.9\linewidth]{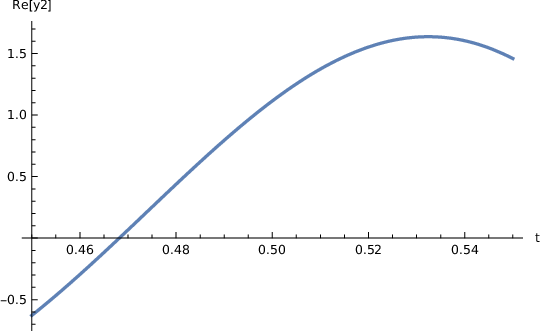}
			\caption{Real part of $y_2$.}
			\label{fig:sub-firstang0}
		\end{subfigure}
		\begin{subfigure}{.5\textwidth}
			\centering
			\includegraphics[width=0.9\linewidth]{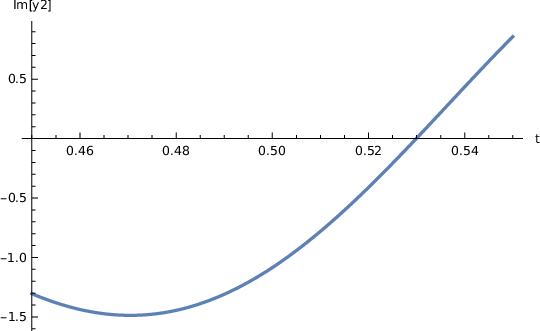}
			\caption{Imaginary part of $y_2$.}
			\label{fig:sub-secondang0}
		\end{subfigure}
		\caption{Plot of $y_2$.}
		\label{fig:DCHcompare}
	\end{figure}
	From now on, we will rename $h(t)$ as $y_1$, $t^{1-\gamma} n(t)$ as $y_2$.
	Now we have to write  $y_1, y_2$ in terms of the two linearly independent solutions, which we name $y_3(1-t)$ and $y_4(1-t)$, the two linearly independent solutions one gets when one expands around $1-t=v$.
	In other words, we want to write 
	\begin{equation}
		y_1(t)= C_1 y_3 (1-t) + C_2 y_4(1-t).
	\end{equation}
	\begin{equation}
		y_2(t)= C_3 y_3 (1-t) + C_4 y_4(1-t).
	\end{equation}
	Note that we had already calculated $y_3$ and $y_4$ above. 
	Using the method given in references \cite {Hatsuda, Noda1, Noda2}, 
	\begin{equation}
		C_1= \frac{W_t [y_1,y_4]}{W_t[y_3,y_4]}, C_2= \frac{W_t [y_1,y_3]}{W_t[y_4,y_3]},  C_3=\frac{W_t [y_2,y_4]}{W_t[y_3,y_4]},  
		C_4=\frac{W_t [y_2,y_3]}{W_t[y_4,y_3]}.
	\end{equation}   
	Here $W_t$ is the Wronskian, where the differentiations are performed with respect to $t$, noting  that $v=1-t$. The calculations were done near the midpoint $t=0.5$.
	\begin{figure}[H]
		\begin{subfigure}{.5\textwidth}
			\centering
			\includegraphics[width=0.9\linewidth]{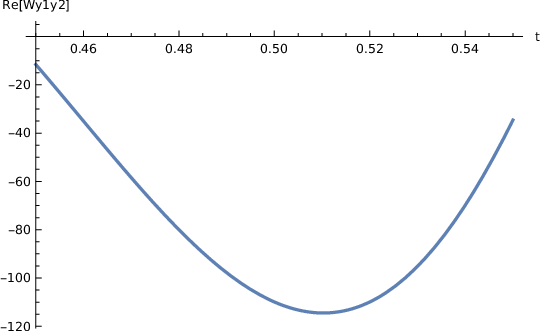}
			\caption{Real part of $W_t[y_1,y_2]$.}
			\label{fig:sub-firstang0}
		\end{subfigure}
		\begin{subfigure}{.5\textwidth}
			\centering
			\includegraphics[width=0.9\linewidth]{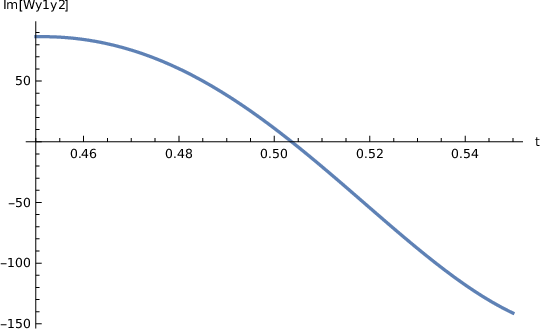}
			\caption{Imaginary part of $W_t[y_1,y_2]$.}
			\label{fig:sub-secondang0}
		\end{subfigure}
		\caption{Plot of $W_t[y_1,y_2]$.}
		\label{fig:DCHcompare}
	\end{figure}
	\begin{figure}[H]
		\begin{subfigure}{.5\textwidth}
			\centering
			\includegraphics[width=0.9\linewidth]{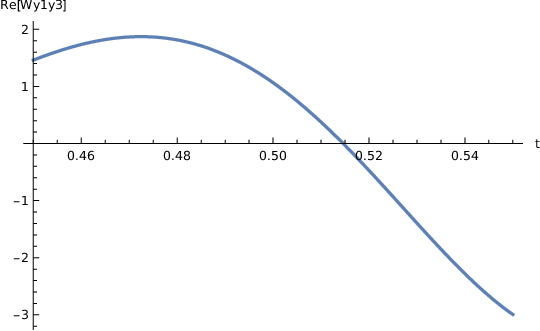}
			\caption{Real part of $W_t[y_1,y_3]$.}
			\label{fig:sub-firstang0}
		\end{subfigure}
		\begin{subfigure}{.5\textwidth}
			\centering
			\includegraphics[width=0.9\linewidth]{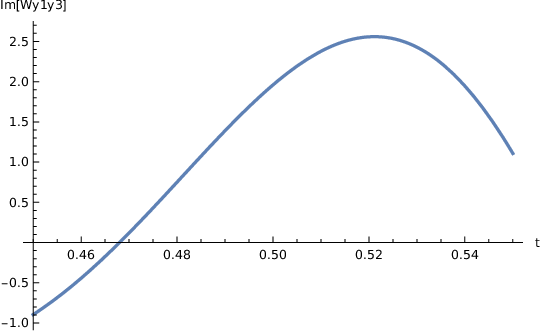}
			\caption{Imaginary part of $W_t[y_1,y_3]$.}
			\label{fig:sub-secondang0}
		\end{subfigure}
		\caption{Plot of $W_t[y_1,y_3]$.}
		\label{fig:DCHcompare}
	\end{figure}
	\begin{figure}[H]
		\begin{subfigure}{.5\textwidth}
			\centering
			\includegraphics[width=0.9\linewidth]{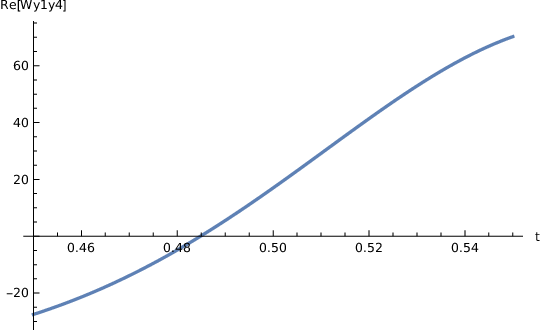}
			\caption{Real part of $W_t[y_1,y_4]$.}
			\label{fig:sub-firstang0}
		\end{subfigure}
		\begin{subfigure}{.5\textwidth}
			\centering
			\includegraphics[width=0.9\linewidth]{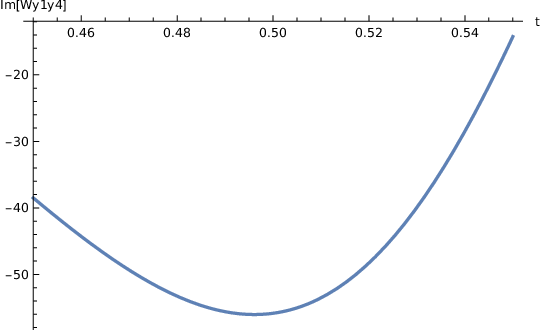}
			\caption{Imaginary part of $W_t[y_1,y_4]$.}
			\label{fig:sub-secondang0}
		\end{subfigure}
		\caption{Plot of $W_t[y_1,y_4]$.}
		\label{fig:DCHcompare}
	\end{figure}
	\begin{figure}[H]
		\begin{subfigure}{.5\textwidth}
			\centering
			\includegraphics[width=0.9\linewidth]{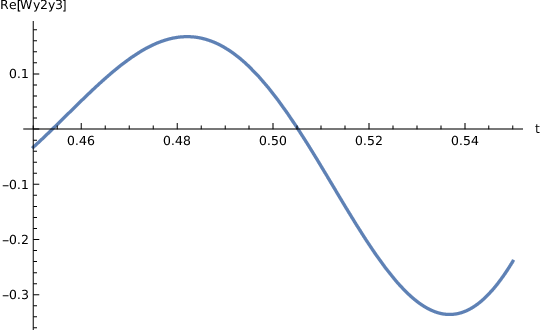}
			\caption{Real part of $W_t[y_2,y_3]$.}
			\label{fig:sub-firstang0}
		\end{subfigure}
		\begin{subfigure}{.5\textwidth}
			\centering
			\includegraphics[width=0.9\linewidth]{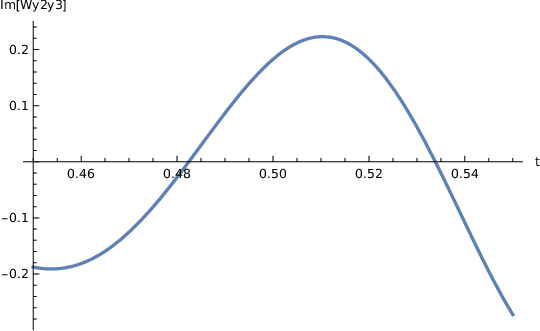}
			\caption{Imaginary part of $W_t[y_2,y_3]$.}
			\label{fig:sub-secondang0}
		\end{subfigure}
		\caption{Plot of $W_t[y_2,y_3]$.}
		\label{fig:DCHcompare}
	\end{figure}
	\begin{figure}[H]
		\begin{subfigure}{.5\textwidth}
			\centering
			\includegraphics[width=0.9\linewidth]{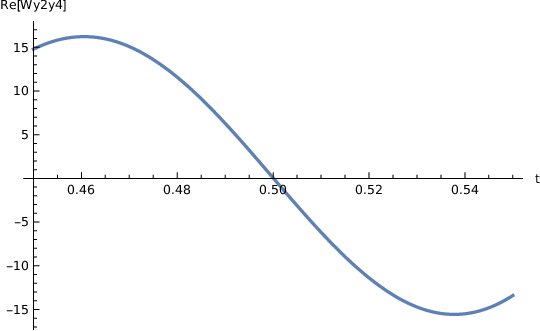}
			\caption{Real part of $W_t[y_2,y_4]$.}
			\label{fig:sub-firstang0}
		\end{subfigure}
		\begin{subfigure}{.5\textwidth}
			\centering
			\includegraphics[width=0.9\linewidth]{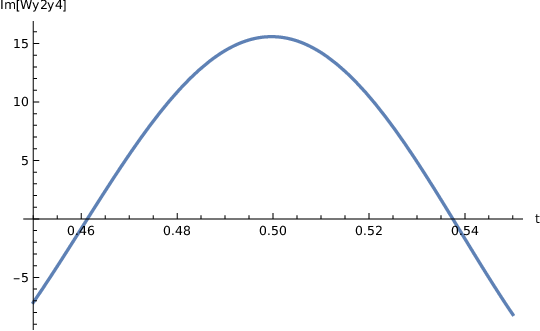}
			\caption{Imaginary part of $W_t[y_2,y_4]$.}
			\label{fig:sub-secondang0}
		\end{subfigure}
		\caption{Plot of $W_t[y_2,y_4]$.}
		\label{fig:DCHcompare}
	\end{figure}
	\begin{figure}[H]
		\begin{subfigure}{.5\textwidth}
			\centering
			\includegraphics[width=0.9\linewidth]{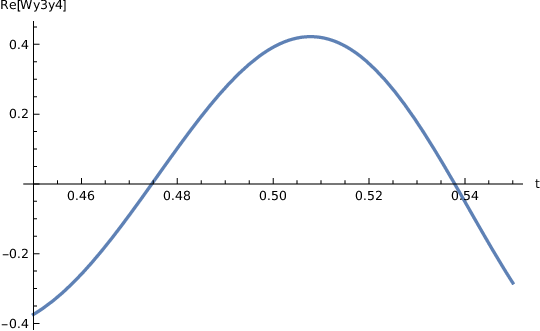}
			\caption{Real part of $W_t[y_3,y_4]$.}
			\label{fig:sub-firstang0}
		\end{subfigure}
		\begin{subfigure}{.5\textwidth}
			\centering
			\includegraphics[width=0.9\linewidth]{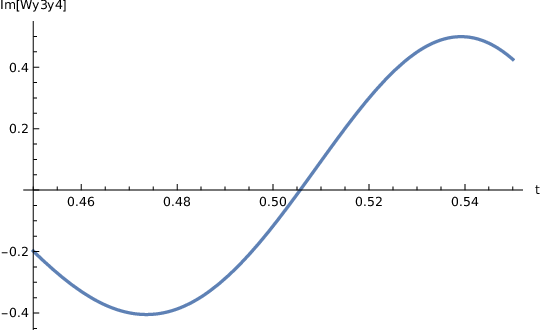}
			\caption{Imaginary part of $W_t[y_3,y_4]$.}
			\label{fig:sub-secondang0}
		\end{subfigure}
		\caption{Plot of $W_t[y_3,y_4]$.}
		\label{fig:DCHcompare}
	\end{figure}
	\begin{figure}[H]
		\begin{subfigure}{.5\textwidth}
			\centering
			\includegraphics[width=0.9\linewidth]{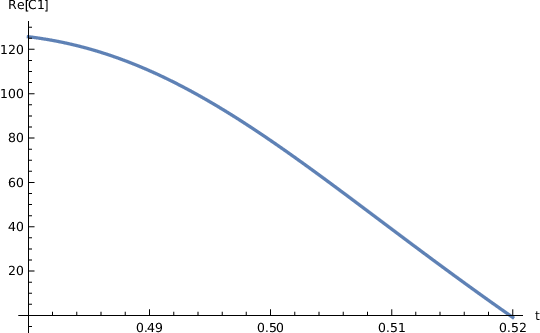}
			\caption{Real part of $C_1$.}
			\label{fig:sub-firstang0}
		\end{subfigure}
		\begin{subfigure}{.5\textwidth}
			\centering
			\includegraphics[width=0.9\linewidth]{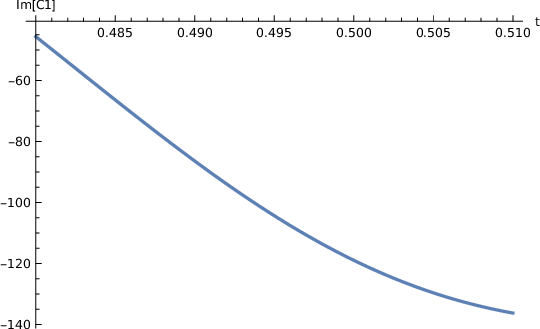}
			\caption{Imaginary part of $C_1$.}
			\label{fig:sub-secondang0}
		\end{subfigure}
		\caption{Plot of $C_1$.}
		\label{fig:DCHcompare}
	\end{figure}
	\begin{figure}[H]
		\begin{subfigure}{.5\textwidth}
			\centering
			\includegraphics[width=0.9\linewidth]{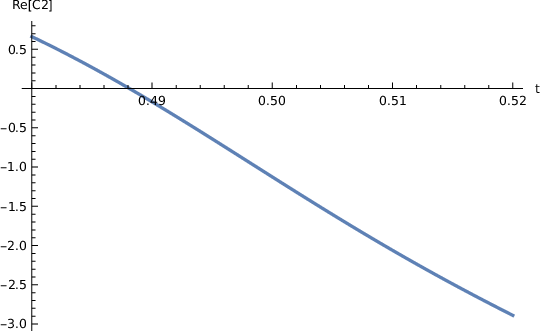}
			\caption{Real part of $C_2$.}
			\label{fig:sub-firstang0}
		\end{subfigure}
		\begin{subfigure}{.5\textwidth}
			\centering
			\includegraphics[width=0.9\linewidth]{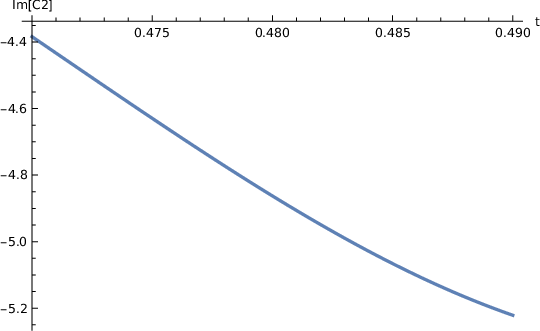}
			\caption{Imaginary part of $C_2$.}
			\label{fig:sub-secondang0}
		\end{subfigure}
		\caption{Plot of $C_2$.}
		\label{fig:DCHcompare}
	\end{figure}
	\begin{figure}[H]
		\begin{subfigure}{.5\textwidth}
			\centering
			\includegraphics[width=0.9\linewidth]{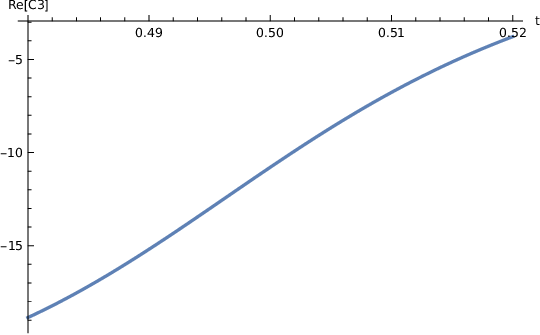}
			\caption{Real part of $C_3$.}
			\label{fig:sub-firstang0}
		\end{subfigure}
		\begin{subfigure}{.5\textwidth}
			\centering
			\includegraphics[width=0.9\linewidth]{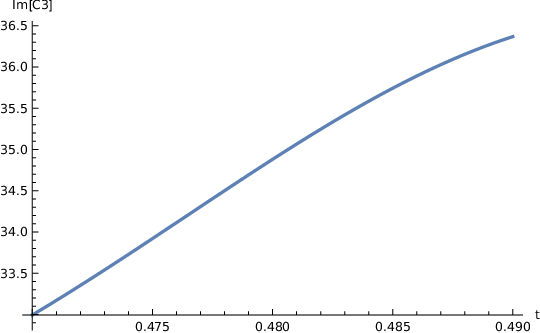}
			\caption{Imaginary part of $C_3$.}
			\label{fig:sub-secondang0}
		\end{subfigure}
		\caption{Plot of $C_3$.}
		\label{fig:DCHcompare}
	\end{figure}
	\begin{figure}[H]
		\begin{subfigure}{.5\textwidth}
			\centering
			\includegraphics[width=0.9\linewidth]{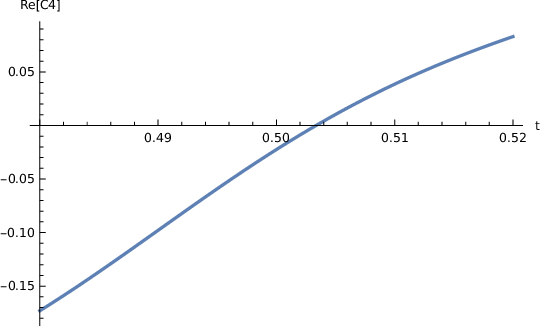}
			\caption{Real part of $C_4$.}
			\label{fig:sub-firstang0}
		\end{subfigure}
		\begin{subfigure}{.5\textwidth}
			\centering
			\includegraphics[width=0.9\linewidth]{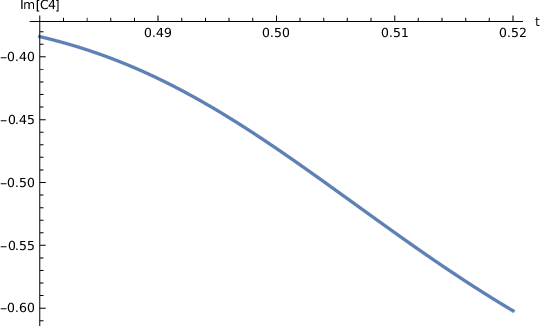}
			\caption{Imaginary part of $C_4$.}
			\label{fig:sub-secondang0}
		\end{subfigure}
		\caption{Plot of $C_4$.}
		\label{fig:DCHcompare}
	\end{figure}
	
	
	One interesting point is that these latter Wronskians do not have constant values when the respective ratios are taken in equation [64] contrary to the claims of our references  \cite {Hatsuda, Noda1, Noda2}. 
	The Wronskians should  become constant when they are divided by $W[y_1,y_2]$, as stated by \cite{Hatsuda, Noda1, Noda2}. But ,these numbers are very very small. Then , if we do not cancel $W[y_1,y_2]$, which divides  both the numerator and the denominator, we get the ratio of two very small numbers. We can not  trust the validity of the result, since these numbers are ratios of two numbers both multiplied by  numbers like $10^{-30}$. When we cancel the $W[y_1,y_2]$ in both terms, before taking the ratios, 
	the figures suggest that the terms $C_i$ given in equation [64] are not constant.  We checked the ratios of \cite {Hatsuda, Noda1, Noda2}. They are indeed constant within certain limits. Also note that the values of the  Wronskians are the same whether we use the form in eq[26] , with the factors multiplying them in the F-homotopic transformation or not. The extra factors cancel in the ratio of two Wronskians.
	
	\section{Conclusion}
	
	Here we put the wave equation for a zero mass scalar particle coupled to the Wahlquist metric to a form to give the Heun solutions in the standard notation and tried to calculate the reflection coefficient for a particle coming from infinity at the origin by two different methods.
	
	\noindent
	Note that this calculation has weak points. We made an approximation by taking $lnx \approx x-1$. This is valid only when $x$ is close to unity. We made a few transformations on the independent variable. We can not guarantee that this approximation is valid in the range we work. We have no choice, though, since we can not solve differential equations whose coefficients are transcendental functions by our codes. 
	
	An interesting point is that we can satisfy an important algebraic constraint, given in \cite{Forsyth, Arscott}, in reducing the first equation to the standard form only if we make this approximation after we calculate $\frac{dU}{dx}$ exactly, and apply this approximation only after the differentiation is done.  
	
	Also note that our coefficients, found in the second method use, are not constants,as they should. This may be due to cetain errors in the exact calculations or to some numerical errors made in our calculations.

	\section{Acknowledgement}
	The first version of this 
	work was written to celebrate the 72th birthday of my dear colleague Prof. Tekin Dereli, whom I met close to fifty years ago, in 1972, when he was still studying for his doctoral degree and I was a fresh Ph.D. , while I was visiting M.E.T.U.in Ankara. We are close friends ever since. The first version was published in "J.Phys.Conf.Ser. 2191 (2022) 1, 012015 • Contribution to: DERELI-FS-2021".
	
	\noindent
	
	In the preparation of this work,  I am grateful to Prof. Nadir Ghazanfari for giving us the key paper by Dekar et. al. I am  also grateful to Prof. Barış Yapışkan for important discussions and correcting a vital mistake in my calculations. I also thank to Prof. Reyhan Kaya, Dr. H.T. \"Oz\c{c}elik and especially Prof. Tolga Birkandan for doing all of the numerical work and drawing the figures. Although I insisted to include his name as an author of this version, I could not persuade him.  I also thank Dr. Oğuzhan Ka\c{s}ık\c{c}ı for technical assistance. 
	
	\noindent
	Work of M.H.  is morally supported by Science Academy, Istanbul, Turkey.
	

\begin{thebibliography}{99}
		
		
		\bibitem{Dariescu} Dariescu M A and  Dariescu C 2021 {\it
			Astrophys.Space Sci.} {\bf366}  44
		
		
		\bibitem{Wahlquist1}  Wahlquist H D 1968 {\it Phys. Rev.} {\bf 172}, 01291 
		
		\bibitem{wahl2} Wahlquist H D 1992 {\it    J.Math. Phys.,Erratum }{\bf{33}} 3235 
		
		\bibitem{wahl3} Wahlquist H D 1993, in {\it{Rotating Objects and Relativistic Physics}}, ed F.J.Chinea and L.M. Gonzalez-Romero, Lecture Notes in Physics Vol. 423 ( Berlin: Springer-Verlag)  p.55 
		
		\bibitem{Schmudzig} Stephani H, Kramer D,  Maccallum, M, Hoenselaer C , Herlt E 2003 {\it{Exact Solutions to Einstein's Field Equations}} 2nd Ed. , Cambridge Monographs on Mathematical Physics, Cambridge University Press , 334 
		
		\bibitem{birkandan}  Birkandan T and Hortacsu M  2021, {\it European Phys.J. } C {\bf 81} 389
		
		\bibitem{Kerr} Kerr R P 1963,{\it Phys. Rev. Letters} {\bf{9}} 237 
		
		\bibitem{Newman} Demianski M and  Newman  E T 1963, {\it Bull. Acad.Polon. Sci.} {\bf 14}, 653 
		
		\bibitem{Jose}  Senovilla J M M 1987,{\it Phys. Lett} A {\bf 123}, 211 
		
		\bibitem{Jose1}  Senovilla J M M 1993, {\it{Rotating Objects and Relativistic Physics}}, edited by Chinea F J  and  Gonzalez-Romero L M , Lecture Notes in Physics Vol. 423 ( Berlin: Springer-Verlag ), p.73
		
		\bibitem{Mars} Mars M 2001,{\it Phys. Rev.} D {\bf 63}, 064022 
		
		\bibitem{Kramer} Kramer D 1985, {\it Class. Quantum Grav.} {\bf 2}, L135 ; 1986 {\it Astron. Nachr.} {\bf307} 309 (1986)
		
		\bibitem{Simon}  Simon W 1984 {\it Gen. Relativ. Grav.} {\bf 16}, 465 
		
		\bibitem{Papa}  Papakostas T 1987, {\it J. Math. Phys.} {\bf 29}, 1445 
		
		\bibitem{Bradley}  Bradley M,  Fodor G,  Marklund M and Perjes Z  2000 {\it
			Class.Quant.Grav.} {\bf{17}}  351  
		
		\bibitem{Bradley1}  Bradley M,  Fodor G and   Perjes Z 2000 {\it
			Class.Quant.Grav.} {\bf{17}}  2635 
		
		\bibitem{bradley2}  Bradley M, Eriksson D,  Fodor G and Racz I  2007, {\it
			Phys.Rev.}
		D {\bf{ 75 }}  024013  
		
		\bibitem{Houri1} Hinoue K, Houri T, Rugina C and  Yasui Y 2014, {\it Phys. Rev.} D {\bf90}, 024037 
		
		\bibitem{Houri2} Houri T ,Tanahashi N and   Yasui Y 2020, {\it Class. Quant. Grav.} {\bf 37}075005  
		
		\bibitem{Arscott}   Arscott F M  1995  , {\textit{Heun's Differential Equations}} ed A. Ronveaux
		, Oxford University Press , p.7.
		
		\bibitem{Forsyth}  Forsyth A 1902 {\it{Theory of Differential Equations, Part III}}, Cambridge University Press 
		
		\bibitem{Arscott1} Arscott F M  1995  , {\textit{Heun's Differential Equations}} ed A. Ronveaux
		, Oxford University Press 
		, p. 19 
		
		
		\bibitem{Suzuki} Suzuki H, Takasugi E and   Umetsu H 1998 {\it Prog. Theor. Phys} 
		{\bf 100}, 491 
		
		\bibitem{Suzuki1}  Suzuki H,  Takasugi E and Umetsu H  {\it Prog. Theor. Phys.} {\bf 103}, 723 
		
		\bibitem{Vieira}  Vieira H S and  Kokkotas, K D 2021 {\it Phys.Rev.} D {\bf 104}   024035  
		\bibitem{Hatsuda} Hatsuda Y 2020 {\it{ Class.Quant.Grav.}} {\bf 38} 2, 025015 • e-Print: 2006.08957
		
		\bibitem{Noda1} Hayato Motohashi, Sousuke Noda 2022 {\it{PTEP}} {\bf 5} 059202 (editorial note), 2021 {\it {PTEP}}  {\bf 8}, 083E03 • e-Print: 2103.10802 
		
		\bibitem{Noda2}Sousuke Noda, Hayato Motohashi 2022
		{\it{Phys.Rev.D }} {\bf 106} 6, 064025 • e-Print: 2206.07721. 
		
		\bibitem{Newman1} Newman E,  Tamburino L and   Unti T 1963,{\it J. Math. Phys.} {\bf 4 }, 915. 
		
		\bibitem{Halilsoy} A-Badavi A and   Halilsoy M 2006,{\it Gen. Relativ. Gravit.} {\bf 38 }, 17629.
		
		\bibitem{Slavyanov} Slavyanov S Y and Lay W 2000 {\it{Special Functions, A Unified Theory Based on Singularities}} Oxford University Press, Oxford. 
		
		\bibitem{Fiziev} Fiziev P P 2010  {\it Class. Quantum Gravity } {\bf{27}}, 135001. 
		
		\bibitem{Hortacsu} Hortacsu M 2018, {\it Adv. High Energy Phys.} {\bf{2018}} 8621573.
		
		\bibitem{Dekar}  Dekar L, Chetouani L,  and  Hammann T F 1998, {\it J. Math. Phys.} {\bf 39} 2551. 
		
		\bibitem{Leaver}   Leaver E W 1986, {\it J. Math. Phys.} {\bf 27}, 1238.
		
		\bibitem{Hatsuda} Hatsuda Y 2020 {\it{ Class.Quant.Grav.}} {\bf 38} 2, 025015 • e-Print: 2006.08957
		
		\bibitem{Noda1} Hayato Motohashi, Sousuke Noda 2022 {\it{PTEP}} {\bf 5} 059202 (editorial note), 2021 {\it {PTEP}}  {\bf 8}, 083E03 • e-Print: 2103.10802 
		
		\bibitem{Noda2}Sousuke Noda, Hayato Motohashi 2022
		{\it{Phys.Rev.D }} {\bf 106} 6, 064025 • e-Print: 2206.07721. 
		
		\ \end{thebibliography}
\end{document}